\documentclass[preprint,aps,tightenlines,nofootinbib,superscriptaddress,preprintnumbers,fleqn,longbibliography]{revtex4-1}

\usepackage{amsmath,graphicx,verbatim,epsfig,amssymb,dsfont}
\usepackage{epstopdf}
\usepackage[latin1]{inputenc}
\usepackage{datetime}
\usepackage[colorlinks=true,linkcolor=blue,citecolor=blue]{hyperref}
\usepackage[usenames,dvipsnames]{color}
\usepackage{setspace}
\usepackage[mathscr]{eucal}

\usepackage{float} 
\usepackage[caption=false]{subfig} 

\newcommand{\Ga}{\Gamma}

\newcommand{\La}{\Lambda}
\newcommand{\nn}{\nonumber}

\newcommand{\bn}{{\bar n}}

\newcommand{\veir}{\varepsilon_{\rm {IR}}}

\newcommand{\be}{\begin{equation}}
\newcommand{\ee}{\end{equation}}
\newcommand{\bea}{\begin{eqnarray}}
\newcommand{\eea}{\end{eqnarray}}
\newcommand{\balign}{\begin{align}}
\newcommand{\ealign}{\end{align}}
\newcommand{\as}{\alpha_s}

\newcommand{\cd}{\!\cdot\!}

\newcommand{\bra}[1]{\left< #1 \right |}
\newcommand{\ket}[1]{\left | #1 \right >}

\newcommand{\sandwich}[3]{\left< #1 \right | #2 \left | #3 \right >}

\newcommand{\bg}{\begin{gather}}
\newcommand{\foma}{\end{gather}}

\newcommand{\noopsort}[1]{}

\newcommand{\vecb}[1]{\mbox{\boldmath $#1$}}
\newcommand{\vecbe}[1]{\mbox{\boldmath ${\scriptstyle #1}$}}
\newcommand{\vecbp}[1]{\mbox{\boldmath $#1_\perp$}}

\def\e{\epsilon}

\def\z{\zeta}

\def\<{\langle}
\def\>{\rangle}

\def\a{\alpha}
\def\b{\beta}
  \def\G{\Ga}
\def\d{\delta}  
\def\s{\sigma}
  
\def\x{\xi}

\def\m{\mu}
\def\n{\nu}

\def\z{\zeta}

\def\({\left(}
\def\[{\left[}
\def\){\right)}
\def\]{\right]}

\def\ln{\hbox{ln}}

\def\le{\left }
\def\ri{\right}

\def\bP{\bar P}

\def\gev{\rm GeV}

\def\lqcd{\La_{\rm QCD}}

\newcommand{\ben}{\begin{eqnarray}}
\newcommand{\een}{\end{eqnarray}}

\newcommand{\bef}{\begin{figure}[htb]\centering}
\newcommand{\eef}{\end{figure}}

\usepackage[normalem]{ulem} 

\begin{document}

\title{
Proper TMD factorization for quarkonia production: $pp\to\eta_{c,b}$ as a study case}

\author{Miguel G. Echevarria}
\email{mgechevarria@pv.infn.it}
\affiliation{Istituto Nazionale di Fisica Nucleare, Sezione di Pavia, via Bassi 6, 27100 Pavia, Italy}

\date{\today}



\begin{abstract}

Quarkonia production in different high-energy processes has recently been proposed in order to probe gluon transverse-momentum-dependent parton distribution and fragmentation functions (TMDs in general).
However, no proper factorization theorems have been derived for the discussed processes, but rather just ansatzs, whose main assumption is the factorization of the two soft mechanisms present in the process: soft-gluon radiation and the formation of the bound state.
In this paper it is pointed out that, at low transverse momentum, these mechanisms are entangled and thus encoded in a new kind of non-perturbative hadronic quantities beyond the TMDs: the TMD shape functions.
This is illustrated by deriving the factorization theorem for the process $pp\to \eta_{c,b}$ at low transverse momentum.




\end{abstract}

\maketitle

\section{Motivation}

Gluons, together with quarks, are the fundamental constituents of nucleons.
They generate almost all their mass and carry about half of their momentum.
However, their three-dimensional (3D) structure is still unknown, as well as their contribution to the nucleon spin.
The 3D structure of hadrons in momentum space is parametrized by the so-called Transverse-Momentum-Dependent Parton Distribution/Fragmentation functions (TMDPDFs/TMDFFs, TMDs in general)~\cite{Angeles-Martinez:2015sea}. 
These are the 3D generalization of the one-dimensional PDFs and FFs, where a dependence on the transverse momentum of the partons is also allowed.
They include as well the correlations between this transverse momentum and the spins of the considered parton and its parent hadron.

Constraining gluon TMDs is a crucial step in our understanding of nucleon 3D structure, and with that also our understanding of confinement in QCD and the structure of ordinary matter in general.
In fact, the study of gluon TMDs in particular, and the gluon content of the nucleons in general, is one of the main motivations that is pushing forward the design of the Electron-Ion Collider in the US~\cite{Accardi:2012qut} and fixed-target experiments at the LHC at CERN~\cite{Hadjidakis:2018ifr,Kikola:2017hnp,Trzeciak:2017csa,Brodsky:2012vg}.

In the last years a huge step forward has been made in the quark TMDs sector, obtaining their proper definition and properties, and their connection with observable cross-sections in terms of robust factorization theorems~\cite{GarciaEchevarria:2011rb,Echevarria:2012pw,Echevarria:2012js,Echevarria:2014rua,Collins:2011zzd}. 
This, together with new higher-order perturbative calculations (see e.g.~\cite{Gutierrez-Reyes:2019rug,Gutierrez-Reyes:2018iod,Echevarria:2016scs,Echevarria:2015byo,Echevarria:2015usa}), has allowed the phenomenological analyses to enter a new precision stage (see e.g. \cite{DAlesio:2014mrz,Echevarria:2014xaa,Bacchetta:2015ora,Bacchetta:2017gcc,Anselmino:2016uie,Scimemi:2017etj,Bertone:2019nxa}).
However, for gluons the situation is very different.
Even if their proper definition is currently known~\cite{Echevarria:2015uaa}, the processes where they can be probed are less clean compared to the ones which are used to access quark TMDs.

All and all, quarkonium production seems the most promising way to probe gluon TMDs.
Indeed, there has been a growing interest lately, with numerous proposals based on tree-level ansatzs for TMD factorization for quarkonium production~\cite{Godbole:2012bx,Boer:2012bt,Godbole:2013bca,Godbole:2014tha,Zhang:2014vmh,Zhang:2015yba,Mukherjee:2015smo,Mukherjee:2016qxa,Mukherjee:2016cjw,Boer:2016bfj,Lansberg:2017tlc,Godbole:2017syo,DAlesio:2017rzj,Rajesh:2018qks,Bacchetta:2018ivt,Lansberg:2017dzg,Kishore:2018ugo,Scarpa:2019fol}
and even several next-to-leading-order (NLO) calculations~\cite{Sun:2012vc,Ma:2012hh,Ma:2014oha,Ma:2015vpt}.

The caveat shared by all these attempts, however, is that all of them assume the decoupling of the two soft mechanisms present in the processes: the soft physics underlying the formation of the quarkonium bound state and the soft gluon resummation.
Indeed, these two soft phenomena cannot be factorized when their relevant scales are comparable, i.e. when $q_T\sim m_Qv$, being $q_T$ the measured transverse momentum and $m_Qv$ the typical momentum of a heavy quark of mass $m_Q$ and velocity $v$ inside the quarkonium state.
For $b\bar b$ states one typically has $v^2\sim 0.1$, while for $c\bar c$ one has $v^2\sim 0.3$, so then $m_bv\sim 1.3~\gev$ and $m_cv\sim 0.7~\gev$.
Roughly speaking, when the transverse momentum is in the non-perturbative region around and below $\lqcd$, then the factorization ansatz made in all previous analyses is not accurate.
And it is precisely this region which is claimed to be sensitive to gluon TMDs and where they can potentially be probed.

In this paper the process $pp\to \eta_{c,b}+X$ is considered as an example of quarkonium production process, and the factorization theorem at low transverse momentum is derived.
It turns out that the cross-section is not given only in terms of gluon TMDs, but there is an additional new non-perturbative hadronic quantity which encodes the two mentioned soft processes together: the TMD shape function.
Using the newly derived factorization theorem, the hard part of the process is obtained at one-loop, which is an essential ingredient for future phenomenological analyses since it allows the resummation of large logarithms at higher logarithmic orders.
Moreover, the obtention of a hard factor free from infrared divergences represents a non-trivial consistency check of the newly derived factorization theorem.

\section{Factorization theorem for $pp\to \eta_{c,b}+X$ at low $q_T$}

The differential cross section for $\eta_{Q}$ ($Q=c,b$) hadro-production is given by
\begin{align}
\label{eq:xsection}
&d\s =
\frac{1}{2s}
\frac{d^3q}{(2\pi)^3 2E_q} \int d^4\x\, e^{-iq\cdot \x}\,
\sum_{X} 
\sandwich{PS_A,\bP S_B}{{\mathcal O}^\dagger(\x)}{X\eta_Q}\,
\sandwich{\eta_Q X}{{\mathcal O}(0)}{PS_A,\bP S_B}
\,,
\end{align}
where $s=(P+\bP)^2$.
The effective operator which mediates this process within an effective theory which combines both soft-collinear effective theory (SCET)~\cite{Bauer:2000ew,Bauer:2000yr,Bauer:2001ct,Bauer:2001yt,Bauer:2002nz,Beneke:2002ph} and non relativistic QCD (NRQCD)~\cite{Bodwin:1994jh} degrees of freedom can be written as~\footnote{A vector $a^\m$ is decomposed as 
$a^\m=\bn\cd a\frac{n^\m}{2}+n\cd a\frac{\bn^\m}{2}+a_\perp^\m=
a^+\frac{n^\m}{2}+a^-\frac{\bn^\m}{2}+a_\perp^\m$, with $n=(1,0,0,1)$, $\bn=(1,0,0,-1)$, $n^2=\bn^2=0$ and $n\cd\bn=2$. We denote $a_T=|\vecb a_\perp|$, i.e. $a_\perp^2=-a_T^2<0$.}
\begin{align}
&{\cal O}(\x) =
-2 q^2 \sum_n
C_H^{(n)}(-q^2;\mu^2)\,
\Big[ \psi^\dagger(\x)\, \G_{\m\n}^{(n)} K_{aa'}^{[1,8]}\, \chi(\x) \Big]\,
\Big[ 
B_{\bn\perp}^{\m,b}(\x)\,
{\mathcal Y}_\bn^{\dagger ba}(\x)\,
{\mathcal Y}_{n}^{a'c}(\x)\,
B_{n\perp}^{\n,c}(\x) 
\Big]
\,,
\end{align}
where the sum runs over the different states $n$ which can contribute to the process.
In the usual spectroscopic notation, $n$ stands for ${}^{2S+1}L_J^{[i]}$, with $S$ the spin, $L$ the angular momentum and $J$ the total angular momentum, and $i=1(8)$ for singlet (octet).
$C_H^{(n)}$ are the spin-independent matching coefficients for each state $n$, which integrate out the hard scale of the process, $q^2=M^2$.
$a,a'b,c$ are the gauge group indexes, while $\G^{(n)}_{\m\n}$ is a matrix which encodes the Lorentz structure of the partonic process for the creation of the state $n$. 
$K_{aa'}$ is a color matrix, with $K_{aa'}^{[1]}=\d_{aa'}$ for color-singlet states and $K_{aa'}^{[8]}=t^A_{aa'}$ for color-octet states.
$\chi$ and $\psi$ are the spinors describing the $Q\bar{Q}$ state.
The $B_{n(\bn)}^{\perp\m}$ operators, which stand for gauge invariant gluon fields, are given by
\begin{align}
&B_{n\perp}^{\m} =
B_{n\perp}^{\m,a} t^a 
=
\frac{1}{g} [W_n^\dagger iD_n^{\perp\m} W_n] 
=
\frac{1}{\bn\cd{\mathcal P}}
i\bn_\a g^{\m}_{\perp\b} W_n^\dagger 
F_n^{\a\b} W_n 
\frac{1}{\bn\cd{\mathcal P}}
i\bn_\a g^\m_{\perp\b} t^a
({\mathcal W}_n^\dagger)^{ab} F_n^{\a\b,b}
\,.
\end{align}
The collinear and soft Wilson lines are the path-ordered exponentials
\begin{align}
W_n(\x) &=
P\exp\le[ ig\int_{-\infty}^{0} ds\, \bn\cd A_n^a(\x+\bn s) t^a\ri]
\,,
\nn\\
Y_n(\x) &=
P\exp\le[ ig\int_{-\infty}^{0} ds\, n\cd A_s^a(\x+n s) t^a\ri]
\,.
\end{align}
Wilson lines with calligraphic typography are in the adjoint representation, i.e., the color generators are given by $(t^a)^{bc}=-if^{abc}$.
In order to guarantee gauge invariance among regular and singular gauges, transverse gauge links need to be added (as described in \cite{Idilbi:2010im,GarciaEchevarria:2011md}).

In the case of $\eta_{Q}$ production, NRQCD formalism dictates that the operator for the state ${}^1S_0^{[1]}$ is the leading one in the power counting in the velocity $v$ \cite{Bodwin:1994jh}.
Thus from now on we will consider only that contribution.
Moreover, the production of color-octet states will potentially spoil TMD factorization due to the presence of uncanceled Glauber gluons~\cite{Collins:2007nk,Collins:2007jp,Rogers:2010dm}.
The Lorentz structure $\G_{\m\n}^{{}^1S_0}$ is fixed by requiring the effective operator to give the same tree-level amplitude as in full QCD for the production of a pseudoscalar $\eta_{Q}$ in the configuration ${}^1S_0^{[1]}$:
\begin{align}
\G_{\m\n}^{{}^1S_0} &= 
\frac{i \pi \as 2\sqrt{2}}{N_c\sqrt{M^5}} 
\epsilon_{\perp \mu\nu}
\,,
\end{align}
where $\epsilon_{\perp}^{\mu\nu} = \epsilon^{\a\b\m\n} n^\a \bn^\b/ (n\cdot\bn)$, with $\e_\perp^{12}=1$.

The effective Lagrangian for this process is given by the combination of both SCET and NRQCD effective Lagrangians.
This means that (anti)collinear modes are decoupled from soft and ultrasoft modes, while the latter are coupled among them (through the NRQCD Lagrangian).
One can thus decompose the final state as the following product of states:
\begin{align}
\label{eq:Xfact}
\ket{X\eta_Q} &= 
\ket{X_n} \otimes \ket{X_\bn} \otimes \ket{X_s\eta_Q}
\,,
\end{align}
where $X_{n,\bn,s}$ are the collinear, anticollinear and soft modes of the unobserved final states.
Notice that the state $\eta_Q$ cannot be decoupled from $X_s$.
A similar decomposition applies to the initial state, considering proton A to be collinear and proton B anticollinear:
\begin{equation}
\label{eq:initialstate}
\ket{P S_A, \bP S_B} = 
\ket{P S_A} \otimes \ket{\bP S_B}
\,.
\end{equation}
 
Using the decompositions in modes~\eqref{eq:Xfact} and~\eqref{eq:initialstate} the cross section is written as
\begin{align}
&d\sigma = 
\frac{1}{2s} \frac{d^3q}{(2\pi)^3 2E_q}\,
4M^4\, H(M^2,\m^2)\, 
\G^*_{\rho\sigma} \G_{\m\n} 
\int d^4 \xi\ e^{-iq\xi} 
\nn\\ 
&\quad 
\times 
\sum_{X_n}\,
\sandwich{P S_A} 
{B_{n\perp}^{\sigma,c'}(\xi)}{X_n}
\sandwich{X_n}
{B_{n\perp}^{\n,c}(0)}
{P S_A}\,
\sum_{X_{\bn}}\, 
\sandwich{\bP S_B}
{B_{\bar{n}\perp}^{\rho,b'}(\xi)}
{X_{\bn}}
\sandwich{X_{\bn}}
{B_{\bn\perp}^{\m,b}(0)}
{\bP S_B}
\nn\\
&\quad 
\times 
\sum_{X_s}
\sandwich{0}{
\Big[
{\mathcal Y}_n^{\dagger c'a'} 
{\mathcal Y}_\bn^{a'b'} 
\chi^\dagger\psi\Big](\x)}
{X_s\eta_Q}\,
\sandwich{\eta_Q X_s}{
\Big[
{\mathcal Y}_\bn^{\dagger ba} 
{\mathcal Y}_n^{ac}
\psi^\dagger\chi\Big](0)}
{0}
\,,
\end{align}
where $H(M^2,\m^2)=|C_H(-q^2;\m^2)|^2$.
The Lorentz structure $\G_{\m\n}\equiv\G_{\m\n}^{{}^1S_0}$ is kept for simplicity.
This result needs to be Taylor expanded in order to extract the leading contribution with a homogenous power counting.
The produced quarkonium is hard with momentum $q \sim M(1,1,\lambda)$, where $\lambda$ is a small parameter parametrizing the relative strength of the momentum components of different modes, $\lambda\sim q_T/M$. 
In the exponent $e^{-iq\xi}$ in~\eqref{eq:xsection} one then has $\xi \sim M(1,1,1/\lambda)$. 
In addition, the scalings of the derivatives of the collinear, anticollinear and soft terms are the same as their respective momentum scalings. 
Given this, the obtained leading term in the Taylor expansion of the cross section is
\begin{align}
&d\sigma = 
\frac{1}{2s} \frac{d^3q}{(2\pi)^3 2E_q}\,
\frac{4M^4\, H(M^2,\m^2)}{(N_c^2-1)^2}\, 
\G^*_{\rho\sigma} \G_{\m\n} 
\int d^4 \xi\ e^{-iq\xi} 
\nn\\ 
&\quad
\times 
\sandwich{P S_A} 
{B_{n\perp}^{\sigma,c}(\x^-,\x_\perp)\, B_{n\perp}^{\n,c}(0)}
{P S_A}\,
\sandwich{\bP S_B} 
{B_{\bn\perp}^{\rho,b}(\x^+,\x_\perp)\, B_{\bn\perp}^{\m,b}(0)}
{\bP S_B}\,
\nn \\
&\quad
\times
\sandwich{0}{
\Big[
{\mathcal Y}_n^{\dagger ca'} 
{\mathcal Y}_\bn^{a'b}
\chi^\dagger\psi\Big](\x_\perp)\,
a_{\eta_Q}^\dagger a_{\eta_Q}\,
\Big[
{\mathcal Y}_\bn^{\dagger ba} 
{\mathcal Y}_n^{ac}
\psi^\dagger\chi\Big](0)}
{0}
+
{\mathcal O}(\lambda)
\,,
\end{align}
where we have also used the fact that (anti)collinear matrix elements are diagonal in color, and the completeness relations
\begin{align}
&\sum_{X_n} \ket{X_n}\bra{X_n} = 1
\,,\quad
\sum_{X_\bn} \ket{X_\bn}\bra{X_\bn} = 1
\,,\nn\\
&\sum_{X_s} \ket{X_s\eta_Q}\bra{X_s\eta_Q} = 
a_{\eta_Q}^\dagger \sum_{X_s} \ket{X_s}\bra{X_s} a_{\eta_Q}
=
a_{\eta_Q}^\dagger a_{\eta_Q}
\,.
\end{align}

Performing standard algebraic manipulations and dropping the suppressed terms, the cross-section can be written as:
\begin{align}
\label{eq:sigma_JJS}
&\frac{d\sigma}{dy d^2q_\perp} =
\frac{4M^4\,H(M^2,\m^2)}{2s M^2 (N_c^2-1)}\, 
\G^*_{\rho\sigma} \G_{\m\n} 
(2\pi)
\int d^2\vecb k_{n\perp} d^2\vecb k_{\bn\perp} d^2\vecb k_{s\perp}\,
\d^{(2)}\le( \vecb q_{\perp}-\vecb k_{n\perp}-\vecb k_{\bn\perp}-\vecb k_{s\perp}\ri)
\nn\\ 
&\quad 
\times
J_n^{(0)\sigma\nu}(x_A,\vecb k_{n\perp},S_A;\m;\d_n)\,
J_{\bn}^{(0)\rho\mu}(x_B,\vecb k_{\bn\perp},S_B;\m;\d_\bn)\,
S_{\eta_Q}^{(0)}\!\Big[{}^1S_0^{[1]}\Big](\vecb k_{s\perp};\m;\d_n,\d_\bn)
\,,
\end{align}
where $x_{A,B}=\sqrt{\tau}\,e^{\pm y}$, $\tau=(M^2+q_T^2)/s$ and $y$ is the rapidity of the produced $\eta_Q$. 
The \emph{pure} collinear matrix elements $J^{(0)}_{n(\bn)}$ and the bare TMD shape function $S_{\mathcal Q}^{(0)}$ (TMDShF from now on) are defined as
\begin{align}
\label{eq:baredefinitions}
&J_n^{(0)\m\n} = 
\frac{x_A P^+}{2} \int\frac{d\x^-d^2\vecb \x_\perp}{(2\pi)^3}\,
e^{-i\le(\frac{1}{2}x_A \x^-P^+-\vecbe \x_\perp\cd\vecbe k_{n\perp}\ri)}\,
\sandwich{PS_A}
{B_{n\perp}^{\m,a}(\x^-,\vecb \x_\perp)\,
B_{n\perp}^{\n,a}(0)}{PS_A}
\,,\nn\\
&J_\bn^{(0)\m\n} = 
\frac{x_B \bP^-}{2} \int\frac{d\x^+d^2\vecb \x_\perp}{(2\pi)^3}\,
e^{-i\le(\frac{1}{2}x_B \x^+\bP^--\vecbe \x_\perp\cd\vecbe k_{\bn\perp}\ri)}\,
\sandwich{\bP S_B}
{B_{\bn\perp}^{\m,a}(\x^+,\vecb \x_\perp)\,
B_{\bn\perp}^{\n,a}(0)}{\bP S_B}
\,,\nn\\
&S_{\eta_Q}^{(0)}\!\Big[{}^1S_0^{[1]}\Big] = 
\frac{1}{N_c^2-1}\,
\int\frac{d^2\vecb \x_\perp}{(2\pi)^2}\,
e^{i\vecbe \x_\perp\cd\vecbe k_{s\perp}}\,
\sandwich{0}{
\Big[
{\mathcal Y}_n^{\dagger ab} 
{\mathcal Y}_\bn^{bc}
\chi^\dagger\psi\Big](\vecb \x_\perp)\,
a_{\eta_Q}^\dagger a_{\eta_Q}\,
\Big[
{\mathcal Y}_\bn^{\dagger cd} 
{\mathcal Y}_n^{da}
\psi^\dagger\chi\Big](0)}
{0}
\,.
\end{align}

Notice that the spurious contribution of the soft momentum modes in the naively calculated collinear matrix elements, denoted $J_{n(\bn)}$ (the so-called ``zero-bin'' in the SCET nomenclature), should be subtracted, in order to avoid their double counting.

Both the collinear matrix elements and the bare TMDShF in \eqref{eq:sigma_JJS} have been written with a dependence on $\d_{n(\bn)}$, which stand for generic rapidity regulators.
These divergences cancel in the full combination of the three matrix elements.
However a different soft function needs to be invoked in order to properly define the gluon TMDs:
\begin{align}
&S = 
\frac{1}{N_c^2-1}\,
\int\frac{d^2\vecb \x_\perp}{(2\pi)^2}\,
e^{i\vecbe \x_\perp\cd\vecbe k_{s\perp}}\,
\sandwich{0}{
\Big[
{\mathcal Y}_n^{\dagger ab}
{\mathcal Y}_\bn^{bc}
\Big](\vecb \x_\perp)\,
\Big[
{\mathcal Y}_\bn^{\dagger cd}
{\mathcal Y}_n^{da}
\Big](0)}
{0}
\,.
\end{align}
This soft function can be split in rapidity space to all orders in perturbation theory as~\cite{Echevarria:2015byo}
\begin{align}\label{eq:softsplitting}
\tilde{S}(\x_T;\m;\d_n,\d_\bn) &=
\tilde{S}_{-}\le(\x_T;\m;\d_n\ri)
\tilde{S}_{+}\le(\x_T;\m;\d_\bn\ri)
\,.
\end{align}
With these pieces, the gluon TMDPDFs are defined as~\cite{Echevarria:2015uaa}
\begin{align}
\label{eq:tmdsdefinition0}
{\tilde G}_{g/A}^{\m\n}(x_A,\vecb \x_{\perp},S_A;\z_A,\m) &=
\tilde{J}_{n}^{(0)\m\n}(x_A,\vecbp \x,S_A;\m;\d_n)\,
\tilde{S}_-(\x_T;\m;\d_n)
\,,
\nn\\
{\tilde G}_{g/B}^{\m\n}(x_B,\vecb \x_{\perp},S_B;\z_B,\m) &=
\tilde{J}_{\bn}^{(0)\m\n}(x_B,\vecbp \x,S_B;\m;\d_\bn)\,
\tilde{S}_+(\x_T;\m;\d_\bn)
\,,
\end{align}
where $\z_{A,B}$ are auxiliary energy scales which arise when the rapidity divergences are cancelled in each TMD, and the twiddle labels the functions in coordinate space.
The chosen rapidity regulator is arbitrary, however the auxiliary energy scales $\z_A$ and $\z_B$ in the TMDs are bound together by $\z_A\z_B=q^4=M^4$.
We emphasize that gluon TMDs so defined are free from rapidity divergences, i.e., they have well-behaved evolution properties and can be extracted from experimental data.

Given the definitions in \eqref{eq:tmdsdefinition0}, the factorized cross-section for proton-proton collisions at low $q_T$ is finally written as
\begin{align}
\label{eq:xsectionfinal}
&\frac{d\sigma}{dy\,d^2q_\perp} = 
\frac{4M^4\,H(M^2,\m^2)}{2s M^2 (N_c^2-1)}\, 
\G^*_{\rho\sigma} \G_{\m\n} 
(2\pi)
\int d^2\vecb k_{n\perp} d^2\vecb k_{\bn\perp} d^2\vecb k_{s\perp}\,
\d^{(2)}\le( \vecb q_{\perp}-\vecb k_{n\perp}-\vecb k_{\bn\perp}-\vecb k_{s\perp}\ri) 
\nn\\
&\quad
\times
G_{g/A}^{\sigma\n}(x_A,\vecb k_{n\perp},S_A;\z_A,\m)\,
G_{g/B}^{\rho\m}(x_B,\vecb k_{\bn\perp},S_B;\z_B,\m)\,
S_{\eta_Q}\!\Big[{}^1S_0^{[1]}\Big](\vecb k_{s\perp};\m)
\,,
\end{align}
where we have defined, for convenience, the TMDShF free from rapidity divergences as
\begin{align}
\label{eq:TMDShF}
{\tilde S}_{\eta_Q}\!\Big[{}^1S_0^{[1]}\Big](\x_T;\m) &=
\frac{\tilde S_{\eta_Q}^{(0)}\!\Big[{}^1S_0^{[1]}\Big](\x_T;\m)}
{\tilde S(\x_T;\m)}
\,.
\end{align}
Given that there are no other vectors available, the TMDShF, as the soft function, depends on the modulus $\x_T$.

The factorization theorem in \eqref{eq:xsectionfinal} is the main result of this letter.
It contains 3 non-perturbative hadronic quantities at low transverse momentum: two gluon TMDPDFs, and the newly defined TMDShF.
Thus, the phenomenological extraction of gluon TMDs from quarkonium production processes is still possible, i.e., a robust factorization theorem can potentially be obtained like in this particular case of $\eta_{c,b}$ hadro-production.
However one also needs to model and extract the involved TMDShFs for the relevant angular/color configurations.

Notice that, while the factorized cross-section in \eqref{eq:xsectionfinal} contains all the (un)polarized gluon TMDs, the TMDShF ${\tilde S}_{\eta_Q}\!\Big[{}^1S_0^{[1]}\Big]$ is spin independent.
In particular, if unpolarized proton collisions are considered, which is relevant e.g. for the LHC, one can parametrize the gluon TMD $G_{g/A}^{\m\n}$ in momentum space as
\begin{equation}
\label{eq:parametrization}
G_{g/A}^{\m\n}(x_A,\vecb k_\perp,S_A;\z_A,\m) \equiv 
\frac{1}{2} \bigg[ - g_\perp^{\mu\n}f_1^g(x_A,k_T^2;\zeta_A,\mu) 
+ \frac{k_T^{\mu\n}}{M_p^2} h_1^{\perp g}(x_A,k_T^2;\zeta_A,\mu) \bigg] \ , 
\end{equation}
where $M_p$ is the mass of the proton and $k_T^{\mu\n}$ is a symmetric traceless tensor of rank 2~\cite{Boer:2016xqr}:
\begin{align}
k_T^{\mu\rho} &= 
k_\perp^\mu k_\perp^\rho + \frac{1}{2} \vecb{k}_\perp^2 g_\perp^{\mu\rho} 
\,.
\end{align}
The function $f_1^g$ is the TMDPDF for unpolarized gluons in an unpolarized proton, while $h_1^{\perp g}$ parametrizes linearly polarized gluons inside an unpolarized proton. 
The parametrization in position space reads
\begin{equation}
\label{eq:parametrization2}
{\tilde G}_{g/A}^{\mu\n}(x_A,\vecb{b}_\perp,S_A;\zeta_A,\mu) \equiv 
\frac{1}{2} \bigg[ 
-g_\perp^{\mu\n} \tilde{f}_1^g(x_A,b_T^2;\zeta_A,\mu)
- \frac{M_p^2}{2} b_T^{\mu\n} \tilde{h}_1^{\perp g (2)}(x_A,b_T^2;\zeta_A,\mu) \bigg]
\,,
\end{equation}
where the Fourier transform of the functions and their moments follow the conventions in~\cite{Boer:2016xqr}. 

Inserting the decomposition \eqref{eq:parametrization} in \eqref{eq:xsectionfinal}, one obtains the factorized cross section for collisions of unpolarized protons:
\begin{align}
\label{e:sigma_ff_hh}
&\frac{d\sigma}{dy d^2q_\perp} =  
\s_0(\m)
H(M^2,\m^2)\,
\Big[
{\mathcal C}[f_1^g f_1^g S_{\mathcal Q}] -
{\mathcal C}[w_{UU} h_1^{\perp g} h_1^{\perp g} S_{Q\bar Q}]
\Big]
\,,
\end{align}
where $S_{\cal Q}$ stands for $S_{\eta_Q}\!\Big[{}^1S_0^{[1]}\Big]$ and the Born-level cross-section is
\begin{align}
\s_0 &=
\frac{(4\pi\as)^2 \pi}{N_c^2 (N_c^2-1) s M^3}
\,.
\end{align}
The convolutions ${\mathcal C}$ are defined in general as:
\begin{align}
&\mathcal{C}[w\, f\, f\, S_{\mathcal Q}] \equiv 
\int d^{2}\vecb p_{\perp a}\,
d^{2}\vecb p_{\perp b}\,
d^{2}\vecb k_{\perp}\,
\delta^{2}(\vecb p_{\perp a}+\vecb p_{\perp b}+\vecb k_{\perp}-\vecb q_{\perp})
\nn\\
&\quad
\times 
w(p_{\perp a},p_{\perp b})\,
f(x_A,\vecb p_{Ta}^2;\zeta_A,\m)\, 
f(x_B,\vecb p_{Tb}^2;\zeta_B,\m)\,
S_{\mathcal Q}(k_T^2;\m)
\,,
\end{align} 
and the transverse momentum weight $w_{UU}$ for the contribution of linearly polarized gluon TMDs is
\begin{equation}
w_{UU} = \frac{p_{\perp a}^{\mu\nu}\ p_{\perp b\,\mu\nu}}{2 M_p^4}
\,.
\end{equation}
Given this, the two Fourier transforms are
\begin{align}
\label{eq:Cff}
&{\mathcal C}[f_1^g f_1^g S_{\mathcal Q}] =
\int \frac{d^2\vecb{b}_T}{(2\pi)^2}\ e^{i \vecbe{b}_T \cdot \vecbe{q}_T}\,
\tilde{f}_1^g(x_A,b_T^2;\zeta_A,\mu)\, 
\tilde{f}_1^g(x_B,b_T^2;\zeta_B,\mu)\,
S_{\mathcal Q}(b_T^2;\m)
\nn\\
&
=
\frac{1}{2\pi} \int_0^{+\infty} db_T b_T J_0(b_T q_T)\,
\tilde{f}_1^g(x_A,b_T^2;\zeta_A,\mu)\, 
\tilde{f}_1^g(x_B,b_T^2;\zeta_B,\mu)\,
S_{\mathcal Q}(b_T^2;\m)
\,, 
\end{align}
and
\begin{align}
\label{eq:Chh}
&{\mathcal C}[w_{UU} h_1^{\perp g} h_1^{\perp g} S_{\mathcal Q}] =
\frac{M_p^4}{16}
\int \frac{d^2\vecb{b}_T}{(2\pi)^2}\ e^{i \vecbe{b}_T \cdot \vecbe{q}_T}\,
\vecb{b}_T^4\, 
\tilde{h}_1^{\perp g (2)}(x_A,b_T^2;\zeta_A,\mu)\,  
\tilde{h}_1^{\perp g (2)}(x_B,b_T^2;\zeta_B,\mu)\,
S_{\mathcal Q}(b_T^2;\m) 
\nn\\
&
=
\frac{M_p^4}{32\pi} \int_0^{+\infty} db_T\,
\vecb{b}_T^5\, J_0(b_T q_T)\,
\tilde{h}_1^{\perp g (2)}(x_A,b_T^2;\zeta_A,\mu)\, 
\tilde{h}_1^{\perp g (2)}(x_B,b_T^2;\zeta_B,\mu)\,
S_{\mathcal Q}(b_T^2;\m) 
\,.
\end{align} 

\section{Calculation of the hard part at NLO}
The calculation of the hard part of the process not only provides a necessary ingredient to perform the resummation of large logarithms to get more reliable results, but it is also a test of the newly derived factorization theorem.

A necessary condition for the factorized cross-section to be correct, is that it has to exactly reproduce all the infrared physics of the cross-section in full QCD, order by order in perturbation theory.
In other words, the hard factor should turn out to be a finite quantity, just an expansion in $\as$.

It is worth emphasizing that the hard part will only come from the miss-match of virtual diagrams in the full theory and the factorized expression, since by construction it only depends on the hard scale $M$, and diagrams with real gluons will have a dependence on the transverse momentum, which is a lower scale.
Thus one just needs to compute the virtual part of the cross-section in QCD and then subtract the virtual parts of the two gluon TMDs and the TMDShF.

Let us start with the virtual part of the cross-section up to ${\mathcal O}(\as)$ which, after renormalization (i.e. after removing ultraviolet divergences), in coordinate space is \cite{Petrelli:1997ge}
\begin{align}
&\frac{d\s}{\s_0}\Bigg|_v =
\d(1-x_A)\d(1-x_B)
+
\frac{\as}{2\pi}\Bigg[
C_F\frac{\pi^2}{v}
- 2 \Bigg(
\frac{C_A}{\veir^2} + \frac{1}{\veir} 
\Bigg(
\frac{\b_0}{2} + C_A\ln\frac{\m^2}{M^2}
\Bigg)
\nn\\
&\quad
- C_A\ln^2\frac{\m^2}{M^2}
- C_A\frac{\pi^2}{6}
+ 2B_{^1S_0}^{[1]}
\Bigg)
\Bigg]
\d(1-x_A) \d(1-x_B)
\,,
\end{align}
with 
\begin{align}
B_{^1S_0}^{[1]} &=
C_F\Bigg(-5+\frac{\pi^2}{4}\Bigg)
+ C_A \Bigg(1+\frac{5\pi^2}{12}\Bigg)
\,.
\end{align}

The virtual contribution of the renormalized TMDPDF in coordinate space can be obtained e.g. from \cite{Echevarria:2015uaa}:
\begin{align}
&{\tilde f}_1^g\Big|_v =
\d(1-x)
+
\frac{\as}{2\pi}\Bigg[
- \frac{C_A}{\veir^2} - \frac{1}{\veir} 
\Bigg(
\frac{\b_0}{2} + C_A\ln\frac{\m^2}{M^2}
\Bigg)\Bigg]
\d(1-x)
\,.
\end{align}

\begin{figure*}[t!]
\centering
\subfloat[][]{
\includegraphics[width=0.3\textwidth]{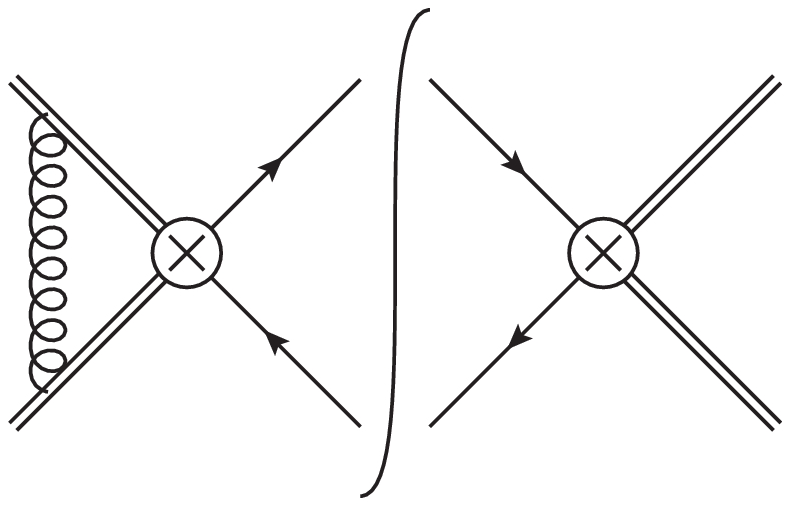}    
\label{fig:ShF-a}}
\qquad
\subfloat[][]{
\includegraphics[width=0.3\textwidth]{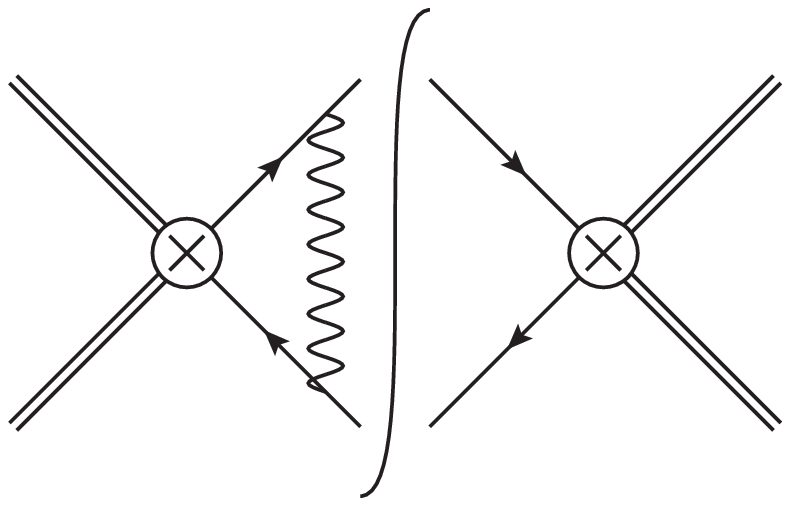}    
\label{fig:ShF-b}}
\caption{\emph{Relevant Feynman diagrams for the virtual part of the TMDShF.
Double solid lines stand for soft Wilson lines, while single lines for heavy quarks.
The gluon in diagram \ref{fig:ShF-a} is soft, while the one in diagram \ref{fig:ShF-b}, denoted with a wavy line, is ultrasoft.
Corresponding crossed diagrams should be added.}}
\label{fig:ShF}
\end{figure*}

The one-loop virtual part of the renormalized TMDShF, defined in \eqref{eq:TMDShF}, is given by the virtual diagrams of the bare TMDShF in figure \ref{fig:ShF} and the ones of the soft function. 
On one hand, diagram \ref{fig:ShF-a} (and its crossed one) gives exactly the same as the corresponding one for the soft function $S$, and thus it is cancelled in ${\tilde S}_{\cal Q}$.
On the other, diagram \ref{fig:ShF-b} (and its crossed one) is analogous to the one found in the long-distance matrix element (LDME) 
$\sandwich{0}{
\chi^\dagger\psi\,
a_{\eta_Q}^\dagger a_{\eta_Q}\,
\psi^\dagger\chi}{0}
$, 
and can thus be obtained e.g. from \cite{Petrelli:1997ge,Jia:2011ah}.
Putting everything together, the result is:
\begin{align}
{\tilde S}_{\mathcal Q}\Big|_{v} &=
1
+
\frac{\as}{2\pi}
C_F
\frac{\pi^2}{v}
\,.
\end{align}

Notice that the heavy-quark self-energy vanishes on the energy shell (see e.g.~\cite{Bodwin:1994jh}).
In addition, there are no interactions at this order between the heavy quarks (soft) and the soft gluons from the soft Wilson lines \cite{Luke:1999kz}.
In fact, the gluon connecting the two soft Wilson lines in diagram \ref{fig:ShF-a} is soft, while the one connecting the heavy quarks in diagram \ref{fig:ShF-b} is ultrasoft.
This is the reason why the authors in \cite{Ma:2012hh,Ma:2015vpt} get to the misleading conclusion that the factorization ansatz they propose for this process is justified, i.e., that the cross-section is given in terms of two (subtracted) gluon TMDs and the local LDME, which they claim is completely factorized from the soft function at low transverse momentum.
Indeed, it turns out that, at one-loop, the virtual part of the TMDShF $S_{\mathcal Q}$ is given by the virtual part of the local LDME.
However this fact does not hold for higher-orders.
The TMDShF at low transverse momentum is a genuine non-perturbative quantity.

Finally, subtracting to the virtual part of the cross-section in full QCD the virtual part of two gluon TMDs and the virtual part of the TMDShF, we obtain the hard part up to ${\mathcal O}(\as)$: 
\begin{align}
H &=
1 +
\frac{\as}{2\pi} \Bigg[
- C_A\ln^2\frac{\m^2}{M^2}
- C_A\frac{\pi^2}{6}
+ 2B_{^1S_0}^{[1]}
\Bigg]
\,,
\end{align}
As expected, this coefficient turns out to be free from infrared divergences, which means that the derived factorization theorem properly reproduces the infrared part of the cross-section in full QCD at one loop.
This constitutes a non-trivial consistency check.
In addition, this coefficient is a necessary ingredient for the resummation of large logarithms at higher orders, allowing for precise phenomenological studies in the near future.

\section{Conclusions}
By applying the effective field theory approach, a proper factorization theorem for $\eta_{c,b}$ hadro-production at low transverse momentum is derived, finding a new kind of non-perturbative hadronic quantity: the TMD shape function (TMDShF).
This matrix element encodes the two soft mechanisms present in the process, the formation of the heavy-quark bound state and the soft-gluon radiation, which were assumed to factorize in all previous works in the literature.

In general, there are as many TMDShFs for a given process as relevant angular/color Fock states within NRCQD power counting.
Simply stated, they could be considered the TMD extensions of the well-known LDMEs.

Quarkonium production processes can thus be used to access gluon TMDs, but the phenomenology is more involved as compared to quark TMDs in, e.g., Drell-Yan or semi-inclusive deep-inelastic scattering processes, since it requires in addition the parametrization of several TMDShFs.

These findings can straightforwardly be applied to other quarkonia production processes, for instance in lepton-hadron collisions (like $ep\to J/\psi$) or electron-positron annihilation (like $e^+e^-\to J/\psi\pi$).
This is left for a future effort.


\section*{Acknowledgements}
The author is supported by the Marie Sk\l odowska-Curie grant \emph{GlueCore} (grant agreement No. 793896).

\bibliographystyle{utphys}
\bibliography{references}

\providecommand{\href}[2]{#2}\begingroup\raggedright\begin{thebibliography}{10}

\bibitem{Angeles-Martinez:2015sea}
R.~Angeles-Martinez {\em et~al.}, ``{Transverse Momentum Dependent (TMD) parton
  distribution functions: status and prospects},''
  \href{http://dx.doi.org/10.5506/APhysPolB.46.2501}{{\em Acta Phys. Polon.}
  {\bfseries B46} no.~12, (2015) 2501--2534},
\href{http://arxiv.org/abs/1507.05267}{{\ttfamily arXiv:1507.05267 [hep-ph]}}.

\bibitem{Accardi:2012qut}
A.~Accardi {\em et~al.}, ``{Electron Ion Collider: The Next QCD Frontier},''
  \href{http://dx.doi.org/10.1140/epja/i2016-16268-9}{{\em Eur. Phys. J.}
  {\bfseries A52} no.~9, (2016) 268},
\href{http://arxiv.org/abs/1212.1701}{{\ttfamily arXiv:1212.1701 [nucl-ex]}}.

\bibitem{Hadjidakis:2018ifr}
C.~Hadjidakis {\em et~al.}, ``{A Fixed-Target Programme at the LHC: Physics
  Case and Projected Performances for Heavy-Ion, Hadron, Spin and Astroparticle
  Studies},''
\href{http://arxiv.org/abs/1807.00603}{{\ttfamily arXiv:1807.00603 [hep-ex]}}.

\bibitem{Kikola:2017hnp}
D.~Kikola, M.~G. Echevarria, C.~Hadjidakis, J.-P. Lansberg, C.~Lorcé,
  L.~Massacrier, C.~M. Quintans, A.~Signori, and B.~Trzeciak, ``{Feasibility
  Studies for Single Transverse-Spin Asymmetry Measurements at a Fixed-Target
  Experiment Using the LHC Proton and Lead Beams (AFTER@LHC)},''
  \href{http://dx.doi.org/10.1007/s00601-017-1299-x}{{\em Few Body Syst.}
  {\bfseries 58} no.~4, (2017) 139},
\href{http://arxiv.org/abs/1702.01546}{{\ttfamily arXiv:1702.01546 [hep-ex]}}.

\bibitem{Trzeciak:2017csa}
B.~Trzeciak, C.~Da~Silva, E.~G. Ferreiro, C.~Hadjidakis, D.~Kikola, J.~P.
  Lansberg, L.~Massacrier, J.~Seixas, A.~Uras, and Z.~Yang, ``{Heavy-ion
  Physics at a Fixed-Target Experiment Using the LHC Proton and Lead Beams
  (AFTER@LHC): Feasibility Studies for Quarkonium and Drell-Yan Production},''
  \href{http://dx.doi.org/10.1007/s00601-017-1308-0}{{\em Few Body Syst.}
  {\bfseries 58} no.~5, (2017) 148},
\href{http://arxiv.org/abs/1703.03726}{{\ttfamily arXiv:1703.03726 [nucl-ex]}}.

\bibitem{Brodsky:2012vg}
S.~J. Brodsky, F.~Fleuret, C.~Hadjidakis, and J.~P. Lansberg, ``{Physics
  Opportunities of a Fixed-Target Experiment using the LHC Beams},''
  \href{http://dx.doi.org/10.1016/j.physrep.2012.10.001}{{\em Phys. Rept.}
  {\bfseries 522} (2013) 239--255},
\href{http://arxiv.org/abs/1202.6585}{{\ttfamily arXiv:1202.6585 [hep-ph]}}.

\bibitem{GarciaEchevarria:2011rb}
M.~G. Echevarria, A.~Idilbi, and I.~Scimemi, ``{Factorization Theorem For
  Drell-Yan At Low $q_T$ And Transverse Momentum Distributions
  On-The-Light-Cone},'' \href{http://dx.doi.org/10.1007/JHEP07(2012)002}{{\em
  JHEP} {\bfseries 07} (2012) 002},
\href{http://arxiv.org/abs/1111.4996}{{\ttfamily arXiv:1111.4996 [hep-ph]}}.

\bibitem{Echevarria:2012pw}
M.~G. Echevarria, A.~Idilbi, A.~Schäfer, and I.~Scimemi, ``{Model-Independent
  Evolution of Transverse Momentum Dependent Distribution Functions (TMDs) at
  NNLL},'' \href{http://dx.doi.org/10.1140/epjc/s10052-013-2636-y}{{\em Eur.
  Phys. J.} {\bfseries C73} no.~12, (2013) 2636},
\href{http://arxiv.org/abs/1208.1281}{{\ttfamily arXiv:1208.1281 [hep-ph]}}.

\bibitem{Echevarria:2012js}
M.~G. Echevarría, A.~Idilbi, and I.~Scimemi, ``{Soft and Collinear
  Factorization and Transverse Momentum Dependent Parton Distribution
  Functions},'' \href{http://dx.doi.org/10.1016/j.physletb.2013.09.003}{{\em
  Phys. Lett.} {\bfseries B726} (2013) 795--801},
\href{http://arxiv.org/abs/1211.1947}{{\ttfamily arXiv:1211.1947 [hep-ph]}}.

\bibitem{Echevarria:2014rua}
M.~G. Echevarria, A.~Idilbi, and I.~Scimemi, ``{Unified treatment of the QCD
  evolution of all (un-)polarized transverse momentum dependent functions:
  Collins function as a study case},''
  \href{http://dx.doi.org/10.1103/PhysRevD.90.014003}{{\em Phys. Rev.}
  {\bfseries D90} no.~1, (2014) 014003},
\href{http://arxiv.org/abs/1402.0869}{{\ttfamily arXiv:1402.0869 [hep-ph]}}.

\bibitem{Collins:2011zzd}
J.~Collins, ``{Foundations of perturbative QCD},''
{\em Camb. Monogr. Part. Phys. Nucl. Phys. Cosmol.} {\bfseries 32} (2011)
  1--624.

\bibitem{Gutierrez-Reyes:2019rug}
D.~Gutierrez-Reyes, S.~Leal-Gomez, I.~Scimemi, and A.~Vladimirov, ``{Linearly
  polarized gluons at next-to-next-to leading order and the Higgs transverse
  momentum distribution},''
\href{http://arxiv.org/abs/1907.03780}{{\ttfamily arXiv:1907.03780 [hep-ph]}}.

\bibitem{Gutierrez-Reyes:2018iod}
D.~Gutierrez-Reyes, I.~Scimemi, and A.~Vladimirov, ``{Transverse momentum
  dependent transversely polarized distributions at
  next-to-next-to-leading-order},''
  \href{http://dx.doi.org/10.1007/JHEP07(2018)172}{{\em JHEP} {\bfseries 07}
  (2018) 172},
\href{http://arxiv.org/abs/1805.07243}{{\ttfamily arXiv:1805.07243 [hep-ph]}}.

\bibitem{Echevarria:2016scs}
M.~G. Echevarria, I.~Scimemi, and A.~Vladimirov, ``{Unpolarized Transverse
  Momentum Dependent Parton Distribution and Fragmentation Functions at
  next-to-next-to-leading order},''
  \href{http://dx.doi.org/10.1007/JHEP09(2016)004}{{\em JHEP} {\bfseries 09}
  (2016) 004},
\href{http://arxiv.org/abs/1604.07869}{{\ttfamily arXiv:1604.07869 [hep-ph]}}.

\bibitem{Echevarria:2015byo}
M.~G. Echevarria, I.~Scimemi, and A.~Vladimirov, ``{Universal transverse
  momentum dependent soft function at NNLO},''
  \href{http://dx.doi.org/10.1103/PhysRevD.93.054004}{{\em Phys. Rev.}
  {\bfseries D93} no.~5, (2016) 054004},
\href{http://arxiv.org/abs/1511.05590}{{\ttfamily arXiv:1511.05590 [hep-ph]}}.

\bibitem{Echevarria:2015usa}
M.~G. Echevarria, I.~Scimemi, and A.~Vladimirov, ``{Transverse momentum
  dependent fragmentation function at next-to--next-to--leading order},''
  \href{http://dx.doi.org/10.1103/PhysRevD.93.011502,
  10.1103/PhysRevD.94.099904}{{\em Phys. Rev.} {\bfseries D93} no.~1, (2016)
  011502}, \href{http://arxiv.org/abs/1509.06392}{{\ttfamily arXiv:1509.06392
  [hep-ph]}}.
[Erratum: Phys. Rev.D94,no.9,099904(2016)].

\bibitem{DAlesio:2014mrz}
U.~D'Alesio, M.~G. Echevarria, S.~Melis, and I.~Scimemi, ``{Non-perturbative
  QCD effects in $q_{T}$ spectra of Drell-Yan and Z-boson production},''
  \href{http://dx.doi.org/10.1007/JHEP11(2014)098}{{\em JHEP} {\bfseries 11}
  (2014) 098},
\href{http://arxiv.org/abs/1407.3311}{{\ttfamily arXiv:1407.3311 [hep-ph]}}.

\bibitem{Echevarria:2014xaa}
M.~G. Echevarria, A.~Idilbi, Z.-B. Kang, and I.~Vitev, ``{QCD Evolution of the
  Sivers Asymmetry},'' \href{http://dx.doi.org/10.1103/PhysRevD.89.074013}{{\em
  Phys. Rev.} {\bfseries D89} (2014) 074013},
\href{http://arxiv.org/abs/1401.5078}{{\ttfamily arXiv:1401.5078 [hep-ph]}}.

\bibitem{Bacchetta:2015ora}
A.~Bacchetta, M.~G. Echevarria, P.~J.~G. Mulders, M.~Radici, and A.~Signori,
  ``{Effects of TMD evolution and partonic flavor on e$^{+}$ e$^{-}$
  annihilation into hadrons},''
  \href{http://dx.doi.org/10.1007/JHEP11(2015)076}{{\em JHEP} {\bfseries 11}
  (2015) 076},
\href{http://arxiv.org/abs/1508.00402}{{\ttfamily arXiv:1508.00402 [hep-ph]}}.

\bibitem{Bacchetta:2017gcc}
A.~Bacchetta, F.~Delcarro, C.~Pisano, M.~Radici, and A.~Signori, ``{Extraction
  of partonic transverse momentum distributions from semi-inclusive
  deep-inelastic scattering, Drell-Yan and Z-boson production},''
  \href{http://dx.doi.org/10.1007/JHEP06(2017)081,
  10.1007/JHEP06(2019)051}{{\em JHEP} {\bfseries 06} (2017) 081},
  \href{http://arxiv.org/abs/1703.10157}{{\ttfamily arXiv:1703.10157
  [hep-ph]}}.
[Erratum: JHEP06,051(2019)].

\bibitem{Anselmino:2016uie}
M.~Anselmino, M.~Boglione, U.~D'Alesio, F.~Murgia, and A.~Prokudin, ``{Study of
  the sign change of the Sivers function from STAR Collaboration W/Z production
  data},'' \href{http://dx.doi.org/10.1007/JHEP04(2017)046}{{\em JHEP}
  {\bfseries 04} (2017) 046},
\href{http://arxiv.org/abs/1612.06413}{{\ttfamily arXiv:1612.06413 [hep-ph]}}.

\bibitem{Scimemi:2017etj}
I.~Scimemi and A.~Vladimirov, ``{Analysis of vector boson production within TMD
  factorization},''
  \href{http://dx.doi.org/10.1140/epjc/s10052-018-5557-y}{{\em Eur. Phys. J.}
  {\bfseries C78} no.~2, (2018) 89},
\href{http://arxiv.org/abs/1706.01473}{{\ttfamily arXiv:1706.01473 [hep-ph]}}.

\bibitem{Bertone:2019nxa}
V.~Bertone, I.~Scimemi, and A.~Vladimirov, ``{Extraction of unpolarized quark
  transverse momentum dependent parton distributions from Drell-Yan/Z-boson
  production},'' \href{http://dx.doi.org/10.1007/JHEP06(2019)028}{{\em JHEP}
  {\bfseries 06} (2019) 028},
\href{http://arxiv.org/abs/1902.08474}{{\ttfamily arXiv:1902.08474 [hep-ph]}}.

\bibitem{Echevarria:2015uaa}
M.~G. Echevarria, T.~Kasemets, P.~J. Mulders, and C.~Pisano, ``{QCD evolution
  of (un)polarized gluon TMDPDFs and the Higgs $q_T$-distribution},''
  \href{http://dx.doi.org/10.1007/JHEP07(2015)158,
  10.1007/JHEP05(2017)073}{{\em JHEP} {\bfseries 07} (2015) 158},
  \href{http://arxiv.org/abs/1502.05354}{{\ttfamily arXiv:1502.05354
  [hep-ph]}}.
[Erratum: JHEP05,073(2017)].

\bibitem{Godbole:2012bx}
R.~M. Godbole, A.~Misra, A.~Mukherjee, and V.~S. Rawoot, ``{Sivers Effect and
  Transverse Single Spin Asymmetry in $e+p^\uparrow \to e+J/\psi+X$},''
  \href{http://dx.doi.org/10.1103/PhysRevD.85.094013}{{\em Phys. Rev.}
  {\bfseries D85} (2012) 094013},
\href{http://arxiv.org/abs/1201.1066}{{\ttfamily arXiv:1201.1066 [hep-ph]}}.

\bibitem{Boer:2012bt}
D.~Boer and C.~Pisano, ``{Polarized gluon studies with charmonium and
  bottomonium at LHCb and AFTER},''
  \href{http://dx.doi.org/10.1103/PhysRevD.86.094007}{{\em Phys. Rev.}
  {\bfseries D86} (2012) 094007},
\href{http://arxiv.org/abs/1208.3642}{{\ttfamily arXiv:1208.3642 [hep-ph]}}.

\bibitem{Godbole:2013bca}
R.~M. Godbole, A.~Misra, A.~Mukherjee, and V.~S. Rawoot, ``{Transverse Single
  Spin Asymmetry in $e+p^\uparrow \to e+J/\psi +X $ and Transverse Momentum
  Dependent Evolution of the Sivers Function},''
  \href{http://dx.doi.org/10.1103/PhysRevD.88.014029}{{\em Phys. Rev.}
  {\bfseries D88} no.~1, (2013) 014029},
\href{http://arxiv.org/abs/1304.2584}{{\ttfamily arXiv:1304.2584 [hep-ph]}}.

\bibitem{Godbole:2014tha}
R.~M. Godbole, A.~Kaushik, A.~Misra, and V.~S. Rawoot, ``{Transverse single
  spin asymmetry in $e+p^\uparrow \to e+J/\psi +X $ and $Q^2$ evolution of
  Sivers function-II},''
  \href{http://dx.doi.org/10.1103/PhysRevD.91.014005}{{\em Phys. Rev.}
  {\bfseries D91} no.~1, (2015) 014005},
\href{http://arxiv.org/abs/1405.3560}{{\ttfamily arXiv:1405.3560 [hep-ph]}}.

\bibitem{Zhang:2014vmh}
G.-P. Zhang, ``{Probing transverse momentum dependent gluon distribution
  functions from hadronic quarkonium pair production},''
  \href{http://dx.doi.org/10.1103/PhysRevD.90.094011}{{\em Phys. Rev.}
  {\bfseries D90} no.~9, (2014) 094011},
\href{http://arxiv.org/abs/1406.5476}{{\ttfamily arXiv:1406.5476 [hep-ph]}}.

\bibitem{Zhang:2015yba}
G.-P. Zhang, ``{Transverse momentum dependent gluon fragmentation functions
  from $J/\psi \ \pi $ production at $e^+ e^-$ colliders},''
  \href{http://dx.doi.org/10.1140/epjc/s10052-015-3723-z}{{\em Eur. Phys. J.}
  {\bfseries C75} no.~10, (2015) 503},
\href{http://arxiv.org/abs/1504.06699}{{\ttfamily arXiv:1504.06699 [hep-ph]}}.

\bibitem{Mukherjee:2015smo}
A.~Mukherjee and S.~Rajesh, ``{Probing Transverse Momentum Dependent Parton
  Distributions in Charmonium and Bottomonium Production},''
  \href{http://dx.doi.org/10.1103/PhysRevD.93.054018}{{\em Phys. Rev.}
  {\bfseries D93} no.~5, (2016) 054018},
\href{http://arxiv.org/abs/1511.04319}{{\ttfamily arXiv:1511.04319 [hep-ph]}}.

\bibitem{Mukherjee:2016qxa}
A.~Mukherjee and S.~Rajesh, ``{$J/\psi $ production in polarized and
  unpolarized ep collision and Sivers and $\cos 2\phi $ asymmetries},''
  \href{http://dx.doi.org/10.1140/epjc/s10052-017-5406-4}{{\em Eur. Phys. J.}
  {\bfseries C77} no.~12, (2017) 854},
\href{http://arxiv.org/abs/1609.05596}{{\ttfamily arXiv:1609.05596 [hep-ph]}}.

\bibitem{Mukherjee:2016cjw}
A.~Mukherjee and S.~Rajesh, ``{Linearly polarized gluons in charmonium and
  bottomonium production in color octet model},''
  \href{http://dx.doi.org/10.1103/PhysRevD.95.034039}{{\em Phys. Rev.}
  {\bfseries D95} no.~3, (2017) 034039},
\href{http://arxiv.org/abs/1611.05974}{{\ttfamily arXiv:1611.05974 [hep-ph]}}.

\bibitem{Boer:2016bfj}
D.~Boer, ``{Gluon TMDs in quarkonium production},''
  \href{http://dx.doi.org/10.1007/s00601-016-1198-6}{{\em Few Body Syst.}
  {\bfseries 58} no.~2, (2017) 32},
\href{http://arxiv.org/abs/1611.06089}{{\ttfamily arXiv:1611.06089 [hep-ph]}}.

\bibitem{Lansberg:2017tlc}
J.-P. Lansberg, C.~Pisano, and M.~Schlegel, ``{Associated production of a
  dilepton and a $\Upsilon(J/\psi)$ at the LHC as a probe of gluon transverse
  momentum dependent distributions},''
  \href{http://dx.doi.org/10.1016/j.nuclphysb.2017.04.011}{{\em Nucl. Phys.}
  {\bfseries B920} (2017) 192--210},
\href{http://arxiv.org/abs/1702.00305}{{\ttfamily arXiv:1702.00305 [hep-ph]}}.

\bibitem{Godbole:2017syo}
R.~M. Godbole, A.~Kaushik, A.~Misra, V.~Rawoot, and B.~Sonawane, ``{Transverse
  single spin asymmetry in $p+p^\uparrow \rightarrow J/\psi +X$},''
  \href{http://dx.doi.org/10.1103/PhysRevD.96.096025}{{\em Phys. Rev.}
  {\bfseries D96} no.~9, (2017) 096025},
\href{http://arxiv.org/abs/1703.01991}{{\ttfamily arXiv:1703.01991 [hep-ph]}}.

\bibitem{DAlesio:2017rzj}
U.~D'Alesio, F.~Murgia, C.~Pisano, and P.~Taels, ``{Probing the gluon Sivers
  function in $p^\uparrow p\to J/\psi\,X$ and $p^\uparrow p \to D\,X$},''
  \href{http://dx.doi.org/10.1103/PhysRevD.96.036011}{{\em Phys. Rev.}
  {\bfseries D96} no.~3, (2017) 036011},
\href{http://arxiv.org/abs/1705.04169}{{\ttfamily arXiv:1705.04169 [hep-ph]}}.

\bibitem{Rajesh:2018qks}
S.~Rajesh, R.~Kishore, and A.~Mukherjee, ``{Sivers effect in Inelastic $J/\psi$
  Photoproduction in $ep^\uparrow$ Collision in Color Octet Model},''
  \href{http://dx.doi.org/10.1103/PhysRevD.98.014007}{{\em Phys. Rev.}
  {\bfseries D98} no.~1, (2018) 014007},
\href{http://arxiv.org/abs/1802.10359}{{\ttfamily arXiv:1802.10359 [hep-ph]}}.

\bibitem{Bacchetta:2018ivt}
A.~Bacchetta, D.~Boer, C.~Pisano, and P.~Taels, ``{Gluon TMDs and NRQCD matrix
  elements in $J/\psi$ production at an EIC},''
\href{http://arxiv.org/abs/1809.02056}{{\ttfamily arXiv:1809.02056 [hep-ph]}}.

\bibitem{Lansberg:2017dzg}
J.-P. Lansberg, C.~Pisano, F.~Scarpa, and M.~Schlegel, ``{Pinning down the
  linearly-polarised gluons inside unpolarised protons using quarkonium-pair
  production at the LHC},''
  \href{http://dx.doi.org/10.1016/j.physletb.2018.08.004,
  10.1016/j.physletb.2019.01.057}{{\em Phys. Lett.} {\bfseries B784} (2018)
  217--222}, \href{http://arxiv.org/abs/1710.01684}{{\ttfamily arXiv:1710.01684
  [hep-ph]}}.
[Erratum: Phys. Lett.B791,420(2019)].

\bibitem{Kishore:2018ugo}
R.~Kishore and A.~Mukherjee, ``{Accessing linearly polarized gluon distribution
  in $J/\psi$ production at the electron-ion collider},''
  \href{http://dx.doi.org/10.1103/PhysRevD.99.054012}{{\em Phys. Rev.}
  {\bfseries D99} no.~5, (2019) 054012},
\href{http://arxiv.org/abs/1811.07495}{{\ttfamily arXiv:1811.07495 [hep-ph]}}.

\bibitem{Scarpa:2019fol}
F.~Scarpa, D.~Boer, M.~G. Echevarria, J.-P. Lansberg, C.~Pisano, and
  M.~Schlegel, ``{Studies of gluon TMDs and their evolution using
  quarkonium-pair production at the LHC},''
\href{http://arxiv.org/abs/1909.05769}{{\ttfamily arXiv:1909.05769 [hep-ph]}}.

\bibitem{Sun:2012vc}
P.~Sun, C.~P. Yuan, and F.~Yuan, ``{Heavy Quarkonium Production at Low Pt in
  NRQCD with Soft Gluon Resummation},''
  \href{http://dx.doi.org/10.1103/PhysRevD.88.054008}{{\em Phys. Rev.}
  {\bfseries D88} (2013) 054008},
\href{http://arxiv.org/abs/1210.3432}{{\ttfamily arXiv:1210.3432 [hep-ph]}}.

\bibitem{Ma:2012hh}
J.~P. Ma, J.~X. Wang, and S.~Zhao, ``{Transverse momentum dependent
  factorization for quarkonium production at low transverse momentum},''
  \href{http://dx.doi.org/10.1103/PhysRevD.88.014027}{{\em Phys. Rev.}
  {\bfseries D88} no.~1, (2013) 014027},
\href{http://arxiv.org/abs/1211.7144}{{\ttfamily arXiv:1211.7144 [hep-ph]}}.

\bibitem{Ma:2014oha}
J.~P. Ma, J.~X. Wang, and S.~Zhao, ``{Breakdown of QCD Factorization for P-Wave
  Quarkonium Production at Low Transverse Momentum},''
  \href{http://dx.doi.org/10.1016/j.physletb.2014.08.033}{{\em Phys. Lett.}
  {\bfseries B737} (2014) 103--108},
\href{http://arxiv.org/abs/1405.3373}{{\ttfamily arXiv:1405.3373 [hep-ph]}}.

\bibitem{Ma:2015vpt}
J.~P. Ma and C.~Wang, ``{QCD factorization for quarkonium production in hadron
  collisions at low transverse momentum},''
  \href{http://dx.doi.org/10.1103/PhysRevD.93.014025}{{\em Phys. Rev.}
  {\bfseries D93} no.~1, (2016) 014025},
\href{http://arxiv.org/abs/1509.04421}{{\ttfamily arXiv:1509.04421 [hep-ph]}}.

\bibitem{Bauer:2000ew}
C.~W. Bauer, S.~Fleming, and M.~E. Luke, ``{Summing Sudakov logarithms in B
  ---> X(s gamma) in effective field theory},''
  \href{http://dx.doi.org/10.1103/PhysRevD.63.014006}{{\em Phys. Rev.}
  {\bfseries D63} (2000) 014006},
\href{http://arxiv.org/abs/hep-ph/0005275}{{\ttfamily arXiv:hep-ph/0005275
  [hep-ph]}}.

\bibitem{Bauer:2000yr}
C.~W. Bauer, S.~Fleming, D.~Pirjol, and I.~W. Stewart, ``{An Effective field
  theory for collinear and soft gluons: Heavy to light decays},''
  \href{http://dx.doi.org/10.1103/PhysRevD.63.114020}{{\em Phys. Rev.}
  {\bfseries D63} (2001) 114020},
\href{http://arxiv.org/abs/hep-ph/0011336}{{\ttfamily arXiv:hep-ph/0011336
  [hep-ph]}}.

\bibitem{Bauer:2001ct}
C.~W. Bauer and I.~W. Stewart, ``{Invariant operators in collinear effective
  theory},'' \href{http://dx.doi.org/10.1016/S0370-2693(01)00902-9}{{\em Phys.
  Lett.} {\bfseries B516} (2001) 134--142},
\href{http://arxiv.org/abs/hep-ph/0107001}{{\ttfamily arXiv:hep-ph/0107001
  [hep-ph]}}.

\bibitem{Bauer:2001yt}
C.~W. Bauer, D.~Pirjol, and I.~W. Stewart, ``{Soft collinear factorization in
  effective field theory},''
  \href{http://dx.doi.org/10.1103/PhysRevD.65.054022}{{\em Phys. Rev.}
  {\bfseries D65} (2002) 054022},
\href{http://arxiv.org/abs/hep-ph/0109045}{{\ttfamily arXiv:hep-ph/0109045
  [hep-ph]}}.

\bibitem{Bauer:2002nz}
C.~W. Bauer, S.~Fleming, D.~Pirjol, I.~Z. Rothstein, and I.~W. Stewart, ``{Hard
  scattering factorization from effective field theory},''
  \href{http://dx.doi.org/10.1103/PhysRevD.66.014017}{{\em Phys. Rev.}
  {\bfseries D66} (2002) 014017},
\href{http://arxiv.org/abs/hep-ph/0202088}{{\ttfamily arXiv:hep-ph/0202088
  [hep-ph]}}.

\bibitem{Beneke:2002ph}
M.~Beneke, A.~P. Chapovsky, M.~Diehl, and T.~Feldmann, ``{Soft collinear
  effective theory and heavy to light currents beyond leading power},''
  \href{http://dx.doi.org/10.1016/S0550-3213(02)00687-9}{{\em Nucl. Phys.}
  {\bfseries B643} (2002) 431--476},
\href{http://arxiv.org/abs/hep-ph/0206152}{{\ttfamily arXiv:hep-ph/0206152
  [hep-ph]}}.

\bibitem{Bodwin:1994jh}
G.~T. Bodwin, E.~Braaten, and G.~P. Lepage, ``{Rigorous QCD analysis of
  inclusive annihilation and production of heavy quarkonium},''
  \href{http://dx.doi.org/10.1103/PhysRevD.55.5853,
  10.1103/PhysRevD.51.1125}{{\em Phys. Rev.} {\bfseries D51} (1995)
  1125--1171}, \href{http://arxiv.org/abs/hep-ph/9407339}{{\ttfamily
  arXiv:hep-ph/9407339 [hep-ph]}}.
[Erratum: Phys. Rev.D55,5853(1997)].

\bibitem{Idilbi:2010im}
A.~Idilbi and I.~Scimemi, ``{Singular and Regular Gauges in Soft Collinear
  Effective Theory: The Introduction of the New Wilson Line T},''
  \href{http://dx.doi.org/10.1016/j.physletb.2010.11.060}{{\em Phys. Lett.}
  {\bfseries B695} (2011) 463--468},
\href{http://arxiv.org/abs/1009.2776}{{\ttfamily arXiv:1009.2776 [hep-ph]}}.

\bibitem{GarciaEchevarria:2011md}
M.~Garcia-Echevarria, A.~Idilbi, and I.~Scimemi, ``{SCET, Light-Cone Gauge and
  the T-Wilson Lines},''
  \href{http://dx.doi.org/10.1103/PhysRevD.84.011502}{{\em Phys. Rev.}
  {\bfseries D84} (2011) 011502},
\href{http://arxiv.org/abs/1104.0686}{{\ttfamily arXiv:1104.0686 [hep-ph]}}.

\bibitem{Collins:2007nk}
J.~Collins and J.-W. Qiu, ``{$k_{T}$ factorization is violated in production of
  high-transverse-momentum particles in hadron-hadron collisions},''
  \href{http://dx.doi.org/10.1103/PhysRevD.75.114014}{{\em Phys. Rev.}
  {\bfseries D75} (2007) 114014},
\href{http://arxiv.org/abs/0705.2141}{{\ttfamily arXiv:0705.2141 [hep-ph]}}.

\bibitem{Collins:2007jp}
J.~Collins, ``{2-soft-gluon exchange and factorization breaking},''
\href{http://arxiv.org/abs/0708.4410}{{\ttfamily arXiv:0708.4410 [hep-ph]}}.

\bibitem{Rogers:2010dm}
T.~C. Rogers and P.~J. Mulders, ``{No Generalized TMD-Factorization in
  Hadro-Production of High Transverse Momentum Hadrons},''
  \href{http://dx.doi.org/10.1103/PhysRevD.81.094006}{{\em Phys. Rev.}
  {\bfseries D81} (2010) 094006},
\href{http://arxiv.org/abs/1001.2977}{{\ttfamily arXiv:1001.2977 [hep-ph]}}.

\bibitem{Boer:2016xqr}
D.~Boer, S.~Cotogno, T.~van Daal, P.~J. Mulders, A.~Signori, and Y.-J. Zhou,
  ``{Gluon and Wilson loop TMDs for hadrons of spin $\leq$ 1},''
  \href{http://dx.doi.org/10.1007/JHEP10(2016)013}{{\em JHEP} {\bfseries 10}
  (2016) 013},
\href{http://arxiv.org/abs/1607.01654}{{\ttfamily arXiv:1607.01654 [hep-ph]}}.

\bibitem{Petrelli:1997ge}
A.~Petrelli, M.~Cacciari, M.~Greco, F.~Maltoni, and M.~L. Mangano, ``{NLO
  production and decay of quarkonium},''
  \href{http://dx.doi.org/10.1016/S0550-3213(97)00801-8}{{\em Nucl. Phys.}
  {\bfseries B514} (1998) 245--309},
\href{http://arxiv.org/abs/hep-ph/9707223}{{\ttfamily arXiv:hep-ph/9707223
  [hep-ph]}}.

\bibitem{Jia:2011ah}
Y.~Jia, X.-T. Yang, W.-L. Sang, and J.~Xu, ``{$O(\alpha_s v^2)$ correction to
  pseudoscalar quarkonium decay to two photons},''
  \href{http://dx.doi.org/10.1007/JHEP06(2011)097}{{\em JHEP} {\bfseries 06}
  (2011) 097},
\href{http://arxiv.org/abs/1104.1418}{{\ttfamily arXiv:1104.1418 [hep-ph]}}.

\bibitem{Luke:1999kz}
M.~E. Luke, A.~V. Manohar, and I.~Z. Rothstein, ``{Renormalization group
  scaling in nonrelativistic QCD},''
  \href{http://dx.doi.org/10.1103/PhysRevD.61.074025}{{\em Phys. Rev.}
  {\bfseries D61} (2000) 074025},
\href{http://arxiv.org/abs/hep-ph/9910209}{{\ttfamily arXiv:hep-ph/9910209
  [hep-ph]}}.

\end{thebibliography}\endgroup

\end{document}